\documentclass[fleqn,usenatbib]{mnras}

\usepackage{newtxtext,newtxmath}

\usepackage[T1]{fontenc}
\usepackage{ae,aecompl}

\usepackage{graphicx}	
\usepackage{amsmath}	
\usepackage{lipsum}
\usepackage{pdflscape}
\usepackage{afterpage}
\usepackage{multirow}

\newcommand{\chandra}{\textit{Chandra}}
\newcommand{\nustar}{\textit{NuSTAR}}

\newcommand{\athena}{{\it ATHENA}}
\newcommand{\xrism}{{\it XRISM}}
\newcommand{\xmm}{{\it XMM-Newton}}

\title[Modelling RMS Spectra II]{Modelling X-ray RMS spectra II: the ultra-fast outflow of PDS~456}

\author[L. H\"arer et al.]{L. H\"arer,$^{1,2}$\thanks{E-mail: lucia.haerer@fau.de}
M. L. Parker,$^{1,3}$\thanks{E-mail: parkermichael75@gmail.com}
A. Joyce,$^{1}$
Z. Igo,$^1$
W. N. Alston,$^{1}$
F. F\"urst,$^{1}$
\newauthor
A. P. Lobban,$^{1}$ 
G. A. Matzeu,$^1$
and
J. N. Reeves$^{4,5}$
\\
$^{1}$European Space Agency (ESA), European Space Astronomy Centre (ESAC), E-28691 Villanueva de la Ca\~{n}ada, Madrid, Spain\\
$^{2}$Dr. Karl Remeis-Observatory \& ECAP, University of Erlangen-Nuremberg, Sternwartstr. 7, 96049 Bamberg, Germany\\
$^{3}$Institute of Astronomy, Madingley Road, Cambridge, CB3 0HA, UK \\
$^{4}$Department of Physics, Institute for Astrophysics and Computational Sciences, The Catholic University of America, Washington, DC 20064, USA \\
$^{5}$INAF, Osservatorio Astronomico di Brera, Via Bianchi 46 I-23807 Merate (LC), Italy
}

\date{Accepted XXX. Received YYY; in original form ZZZ}

\pubyear{2020}

\begin{document}
\label{firstpage}
\pagerange{\pageref{firstpage}--\pageref{lastpage}}
\maketitle

\begin{abstract}

\noindent We present an improved model for excess variance spectra describing ultra-fast outflows and successfully apply it to the luminous ($L_{\rm bol}\sim10^{47}\mathrm{erg}\,\mathrm{s}^{-1}$) low-redshift ($z=0.184$) quasar PDS~456. The model is able to account well for the broadening of the spike-like features of these outflows in the excess variance spectrum of PDS~456, by considering two effects: a correlation between the outflow velocity and the logarithmic X-ray flux and intrinsic Doppler broadening with $v_\mathrm{int} = 10^4\, \mathrm{km}\,\mathrm{s}^{-1}$. The models were generated by calculating the fractional excess variance of count spectra from a Monte Carlo simulation. We find evidence that the outflow in PDS~456 is structured, i.e., that there exist two or more layers with outflow velocities $0.27\mbox{--}0.30\,c$, $0.41\mbox{--}0.49\,c$, and $0.15\mbox{--}0.20\,c$ for a possible third layer, which agrees well with the literature. We discuss the prospects of generally applicable models for excess variance spectra for detecting ultra-fast outflows and investigating their structure. We provide an estimate for the strength of the correlation between the outflow velocity and the logarithmic X-ray flux and investigate its validity. 

\end{abstract}




\section{Introduction}

Active Galactic Nuclei (AGN) have been known to vary rapidly since the early days of X-ray astronomy \citep[e.g.,][]{Barr86,Lawrence87}. With the launch of more sensitive instruments, most notably the European Photon Imaging Camera (EPIC) detectors on board \xmm, it became possible to study the variability of AGN in multiple narrow energy bands, comparable to the energy resolution of the instruments. The most commonly used tool for this is the fractional excess variance ($F_\mathrm{var}$) spectrum, which quantifies the amount of variance above the expected noise level, divided by the mean count rate, as a function of energy \citep{Edelson02, Vaughan03_variability}.

Variance spectra show the energy dependence of variability, so they can be used to try and understand the spectral components responsible for the variability. For example, a common result is that the soft X-ray band is less variable than the hard band, which implies that the commonly observed soft excess in AGN is less variable than the power-law continuum \citep[e.g.,][]{Vaughan04, Parker15_pcasample,Igo20}.

A major limitation of studies using variance spectra is the lack of quantitative analysis. This restricts the results that can be obtained, and makes comparison between different spectra difficult. The typical analysis of such a spectrum is a qualitative examination by eye, and a more sophisticated analysis will include comparison with a simulated spectrum based on the model fit to the count spectrum \citep[e.g.,][]{Vaughan04, Matzeu16, Yamasaki16, Mallick17, Alston20}.

A more robust approach is to construct generally applicable models of spectral variability for different processes, which can be combined with each other, analogous to the models used to fit normal count spectra. \citet{Parker20} constructed some basic models for AGN variability\footnote{publicly available from \url{www.michaelparker.space}} and applied them to the complex variance spectrum of IRAS~13224-3809. These models gave a good description of the data, and allowed the different physical processes contributing to the variance to be quantified.

Recently, \citet{Parker17_irasvariability, Parker18_pds456} showed that ultra-fast outflows \citep[UFOs; e.g.][]{Tombesi10} produce characteristic `spikes' of enhanced variability in a variance spectrum. This is due to their response to the continuum, where the absorption lines from the UFO disappear as the X-ray flux rises \citep{Parker17_nature,Pinto18,Parker18_pds456,Igo20}. This means that the flux in the energy bands affected by absorption rises more, leading to a higher variability amplitude in these bands. When plotted as a variance spectrum, the absorption lines therefore appear as positive spikes in the data. 

There are two main uses for this. Firstly, it can function as a method for UFO detection without complex spectral analysis, and without the same systematic effects that plague conventional spectroscopic detection. \citet{Igo20} used this method with a sample of bright, variable AGN, and found evidence for UFOs in 30--60\% of sources studied, consistent with the results from conventional spectroscopy. Secondly, by modelling the variance spectra of UFOs we can probe the physical properties of the outflow, measuring properties that cannot be measured from a static count spectrum.

A 
relation between the X-ray flux of an AGN and the velocity of its UFO was discovered by \citet{Matzeu17} in PDS~456 and \citet{Pinto18} in IRAS~13224-3809, which may 
result from a radiatively driven wind. So far, the effect of this velocity relation on variance spectra has not been investigated.

PDS\,456 was discovered serendipitously during the Pico del Dias Survey (PDS) by \citet{Torres97} at $z=0.184$ and it is considered the most luminous ($L_{\rm bol}\sim10^{47}\,\rm erg\,s^{-1}$; \citealt{Simpson99,Reeves00}) radio-quiet quasar in the local Universe. Such high luminosity is more typical of sources at red shift $z=2$--$3$, the peak of quasar activity, when AGN feedback played an important role in the evolution of galaxies \citep{DiMatteo05}. 

During a $\sim40\,\rm ks$ \textit{XMM-Newton} snapshot in 2001, \citet{Reeves03} detected a deep absorption trough above $7\,\rm keV$ likely corresponding to highly ionized iron with an outflow velocity of $v_{\rm out}\gtrsim0.10c$. After nearly 20 years of additional studies, the high velocity outflow detected in the Fe~K band is well established \citep[e.g.,][]{Reeves09,Nardini15,Matzeu17,Parker18_pds456,Boissay19} and ranging between $v_{\rm out}\sim0.25$--$0.35c$. A lower ionization counterpart was detected in the soft X-ray band by \citet{Reeves16} in a comprehensive archival (2001--2014) \textit{XMM-Newton} Reflection Grating Spectrometer (RGS) analysis, with a velocity of $0.1$--$0.2c$. 

A faster relativistic disk-wind component, approaching $v_{\rm out}\sim0.5c$, was also observed in the Fe~K band with \nustar~in 2017 \citep{Reeves18_pds456}. This additional feature likely arises from a sub-structure corresponding to the more ionized and innermost region (or streamline) of the disk-wind launched from closer to the black hole. Such a simultaneous detection of a multi-phase disk-wind has only been seen in few other sources, e.g., MCG--03--58--007 \citep{Braito18,Matzeu19} and IRAS\,13449$+$2438 \citep{Parker20_iras13449arXiv}.

Because of the large amount of archival data, complex UFO absorption, and strong variability \citep[e.g., ][]{Reeves20}, PDS~456 is an ideal source for UFO variance spectroscopy.
In this paper, we present an improved \textsc{xspec} model for UFO variance, accounting for variability in the velocity of the absorber, and show that this model can explain the variance spectrum of PDS~456.

\section{Observations and Data Reduction}

We use all archival \xmm\ observations of PDS~456 included in the analysis of \citet{Igo20}, with the addition of three observations taken in 2018 and 2019 \citep[obsIDs 0830390101, 0830390201, and 0830390401;][]{Reeves20}. 

In general, we follow the same data reduction procedure as \citet{Igo20}. We use only the high signal EPIC-pn data, which is less affected by background photons at high energies (where UFO features are present). We reduce the data using the \textsc{epproc} tool, part of the \xmm\ Science Analysis Software (SAS, version 18.0.0), with the standard settings. We filter for background flares, and extract source photons from a $20^{\prime\prime}$ circular region centered on the source coordinates, and background photons from a larger ($60^{\prime\prime}$) circular region on the same chip, avoiding the chip gap, regions with high copper background, and other X-ray sources. 

To construct the variance spectrum, we extract 200 lightcurves from logarithmically spaced energy bins, with a time step of 10000s. The total clean exposure is 890~ks. Following \citeauthor{Igo20} we rebin the lightcurves to find an optimum compromise between signal to noise and energy resolution. In this case, a binning the lightcurves by a factor of 2 (100 energy bins, $\Delta E/E = 0.026$) gives a spectrum with clear features and good signal.


\section{Improved UFO model}
\label{sec:models}

The Fe XXVI K$\alpha$ UFO line visible in the RMS spectrum of PDS~456 is much broader than the corresponding feature for IRAS~13224-3809 \citep{Parker18_pds456,Igo20}. One potential reason for this is that the strong velocity variability identified by \citet{Matzeu17}, who showed that the UFO velocity is strongly correlated with the X-ray flux. We therefore construct a model that can take this variability into account.

Following \citet{Parker20}, we use a Monte Carlo approach to generate a table model. For each grid point 1000 spectra are simulated using \textsc{xspec} 12.11.0m \citep{Arnaud96}, whereby one parameter is varied and the other parameters are either correlated with it or fixed. Then, $F_\mathrm{var}$ is obtained by calculating the variance and mean of these spectra for each energy bin.

\citet{Parker20} applied this method to generate a model \textsc{fvar\_ufo} for UFOs and successfully described the variance spectrum of IRAS~13224-3809. In this model, the flux, $F$, is varied by drawing values for $\log F$ from a normal distribution with a fixed variance of 0.2. The logarithm of the mean ionization, $\log\xi$, is correlated with $\log F$ via a multiplicative factor $C_\mathrm{UFO}$. This correlation is motivated by the anticorrelation between the strength of UFO absorption lines and the continuum, which was found in various studies, e.g., \citet{Parker17_nature} and \citet{Pinto18}. This behaviour can be explained by the fact that there is less absorption in a gas with higher $\chi$, because fewer electrons are bound and hence able to absorb X-rays. As, e.g., \citet{Fukumura18} discuss, this picture may be an oversimplification. However, it is consistent with the data and therefore sufficient for the purposes of this model. In addition to $\xi$ and $C_\mathrm{UFO}$, the model accounts for the column density, $n_\mathrm{H}$, of the outflowing gas. The model is based on the \textsc{xabs} count spectral model \citep{Steenbrugge03} from \textsc{spex} \citep{Kaastra96} and is calculated as a multiplicative table model relative to a power law component, i.e, a power law is included for the simulation and divided out later. 

\begin{figure}
    \centering
    \includegraphics[width=\linewidth]{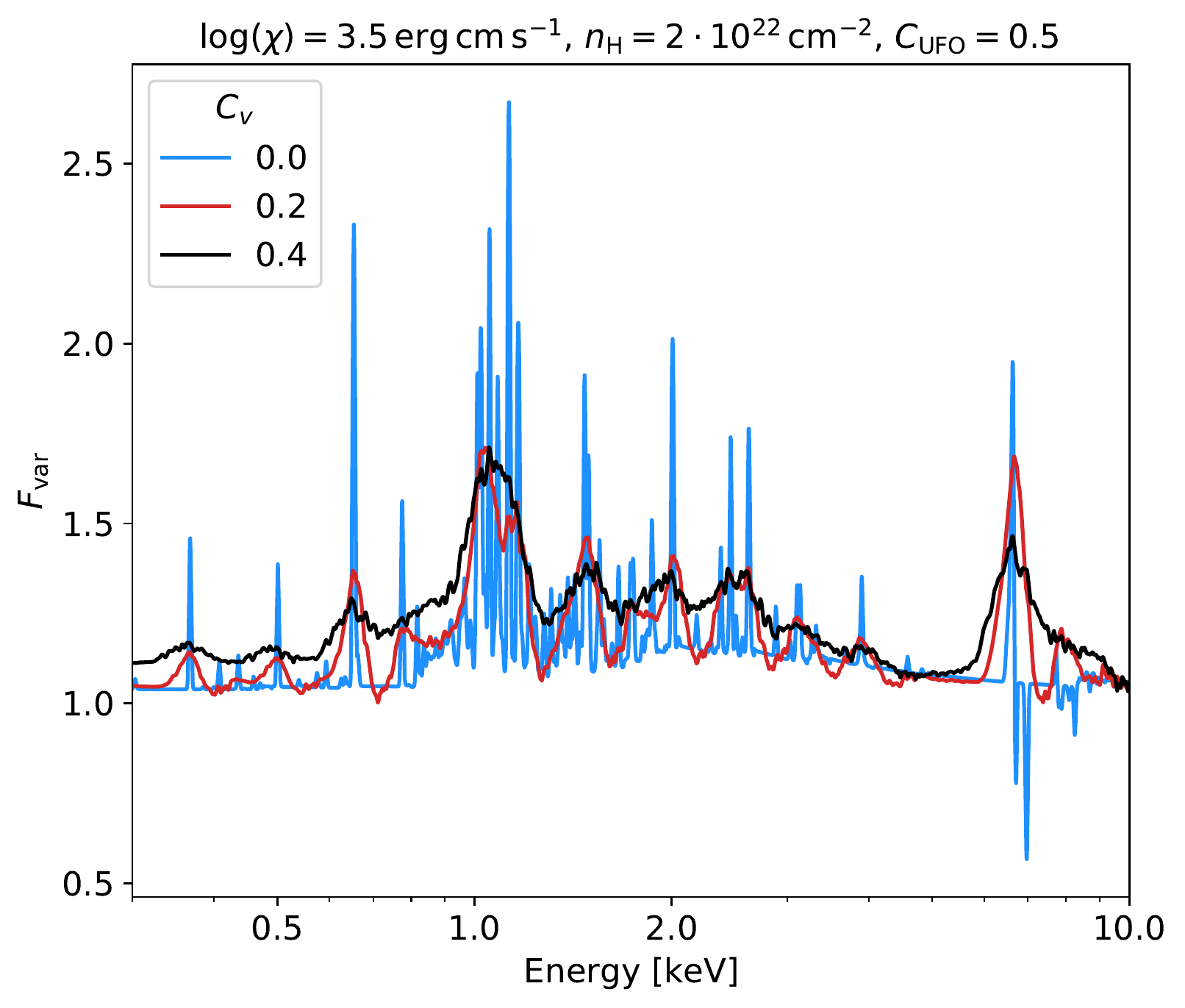}
    \caption{The RMS spectrum as predicted by the improved UFO model, which includes a correlation between the outflow velocity and the logarithm of the flux. Three different values of the correlation parameter, $C_v$, are given. The correlation causes a broadening of the UFO absorption spikes.}
    \label{fig:models}
\end{figure}

We improve upon the \textsc{fvar\_ufo} model by introducing a correlation between the outflow velocity and the logarithm of the X-ray flux (velocity trend hereafter). \citet{Matzeu17} found such a trend in observations of PDS~456. In practice, this means that, due to the change in outflow velocity, the UFO absorption lines in the count spectrum get shifted when the luminosity changes, leading to an increase of variability around the mean line positions. 
For practical reasons, we implement the outflow velocity in terms of the redshift parameter, $z$. Figure~\ref{fig:models} shows the model spectrum for three different correlation strengths, given in terms of the velocity correlation parameter, $C_v$. It can be seen that an increase in $C_v$ causes a broadening of the absorption spikes. This broadening results from the shift in the UFO line positions caused by the velocity trend, as explained above.

\begin{figure}
    \centering
    \includegraphics[width=\linewidth]{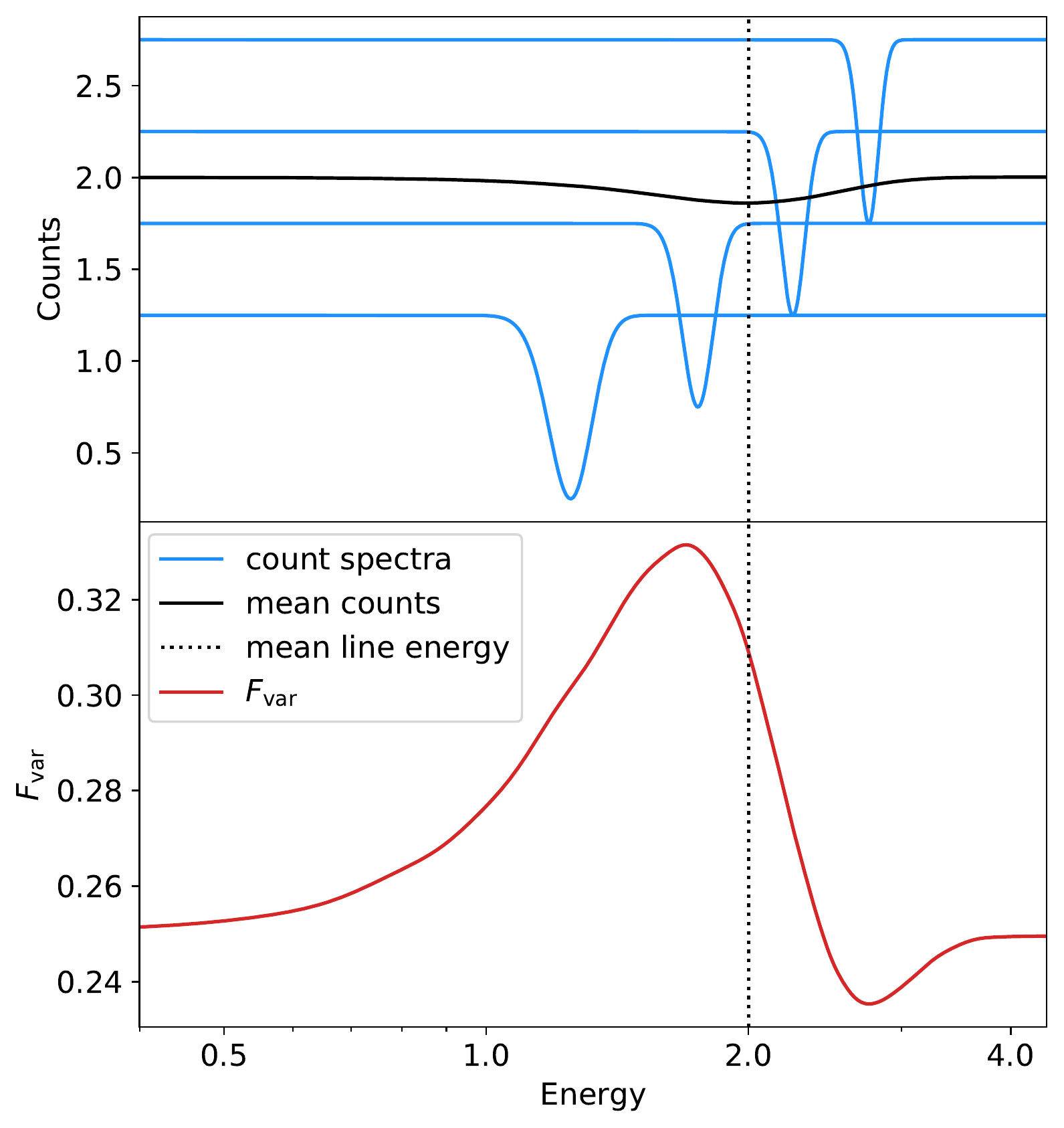}
    \caption{The shape of a spike in the RMS spectrum resulting from a correlation of the absorption line energy and the number of counts, as, e.g., given by the velocity trend in PDS~456. The spike peaks slightly below the mean line energy (\textit{dotted line}) in the count spectrum, because the spectra with high count rate cause a dip in variance at the high energy side of the line. For this example, 10000 spectra with a single Gaussian absorption line of fixed variance of 0.1 and depth of 1 were generated (\textit{blue}). The number of counts was drawn from a normal distribution with mean 2 and variance 0.5 and correlated to the number of counts with a factor of 1. From these spectra, the mean number of counts (\textit{black}) and $F_\mathrm{var}$ (\textit{red}) were calculated. The dotted line indicates the mean line energy. All units are arbitrary.}
    \label{fig:dip}
\end{figure}

In addition, the shape of the spike changes as well, as Fig.~\ref{fig:dip} shows. The maximum is slightly redshifted compared to the mean absorption line energy in the count spectrum, due to a dip at the high energy side of the line. This dip appears because the relative change in count rate is lower if the line is maximally blueshifted. For high blueshift, the count rate is above average, due to the velocity trend. In this case, the drop caused by an absorption line brings it closer to average (see blue and black lines in Fig.~\ref{fig:dip}), decreasing the variance. Note that the strength of the dip depends on $C_v$, because a higher $C_v$ leads to lines which are broader but not as deep, which counteracts the effect. In addition, the strength of the dip depends on the distribution of line energies and on the variation of line shape with flux. In the example in Fig.~\ref{fig:dip}, the line energies are normally distributed and the line broadness and depth are constant. Also note that in Fig.~\ref{fig:models}, the dips are not always clearly visible because spikes blend into each other. 

There is also an additional mechanism that can lead to dips in the variance spectrum. As previously explained, we ascribe the weakening of absorption lines with flux to an increase in ionization. As a result, the abundance of Fe XXVI increases with flux, which counteracts the overall decrease of line strength and decreases the variability at the high end of the Fe~K$\alpha$ line, because its energy is slightly higher for Fe XXVI.

\begin{figure}
    \centering
    \includegraphics[width=\linewidth]{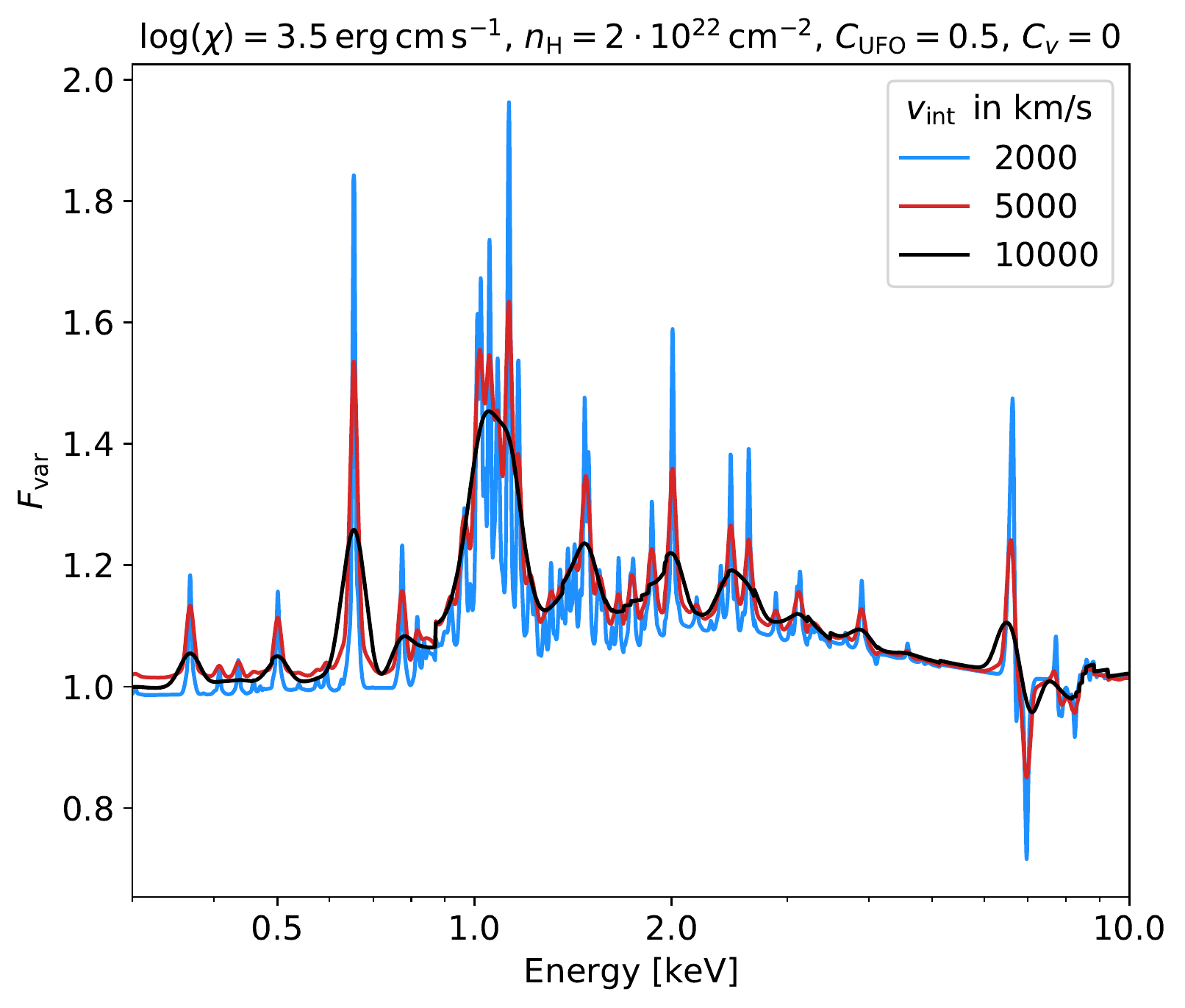}
    \caption{The RMS spectrum for the second version of the improved UFO model, which includes Doppler broadening due the movement of the gas with a velocity $v_\mathrm{int}$. A higher $v_\mathrm{int}$ results in a smoothing and broadening of the spikes. }
    \label{fig:models_broad}
\end{figure}

As a second improvement to the $F_\mathrm{var}$ UFO model besides the introduction of a velocity trend, we take into account the intrinsic velocity of the gas surrounding the AGN, $v_\mathrm{int}$. In count spectra, $v_\mathrm{int}$ results in a Doppler broadening of the absorption lines. Figure~\ref{fig:models_broad} shows the effect of this broadening on our model. As explained above, we consider the $F_\mathrm{var}$ absorption spikes to be due to the change of line depth with ionization. Introducing line broadening in the count spectrum increases the energy range around the center of the line where this effect plays a role and therefore also broadens the spikes. 

In summary, both a velocity trend and $v_\mathrm{int}$ cause a broadening of the absorption spikes in RMS spectra. We therefore have created two versions of the model: \texttt{fvar\_ufo.fits} which takes into account only a velocity trend, and \texttt{fvar\_ufo\_vint.fits} with an additional fixed $v_\mathrm{int}=10^4\,\mathrm{km}\,\mathrm{s}^{-1}$. This value is motivated by \citet{Matzeu16} and \citet{Reeves16}, who both find $v_\mathrm{int} \sim 10^4\,\mathrm{km}\,\mathrm{s}^{-1}$. We use both models in our analysis of PDS~456 and discuss their physical implications and differences in the following sections.      

\section{Results}
\label{sec:results}

Fitting RMS spectra differs in some respects from fitting count spectra. The general considerations are laid out in \cite{Parker20}, who have introduced the method we use for model generation. To account for the resolution of \textit{XMM-Newton's} EPIC-pn, the model is smoothed by an energy-dependent Gaussian (\textsc{gsmooth} in \textsc{xspec}) with a standard deviation of $0.1\,$keV at $6\,$keV and an energy dependence index of 0.165. A 1\% systematic error was added to avoid overprecise fitting at low energies at the expense of higher energies, which is necessary because the models are not yet as precise as count spectral models and the precision of $F_\mathrm{var}$ is significantly higher at lower energies.

\subsection{Fitting spectra of PDS~456}

\begin{figure*}
    \centering
    \includegraphics[width=0.45\linewidth]{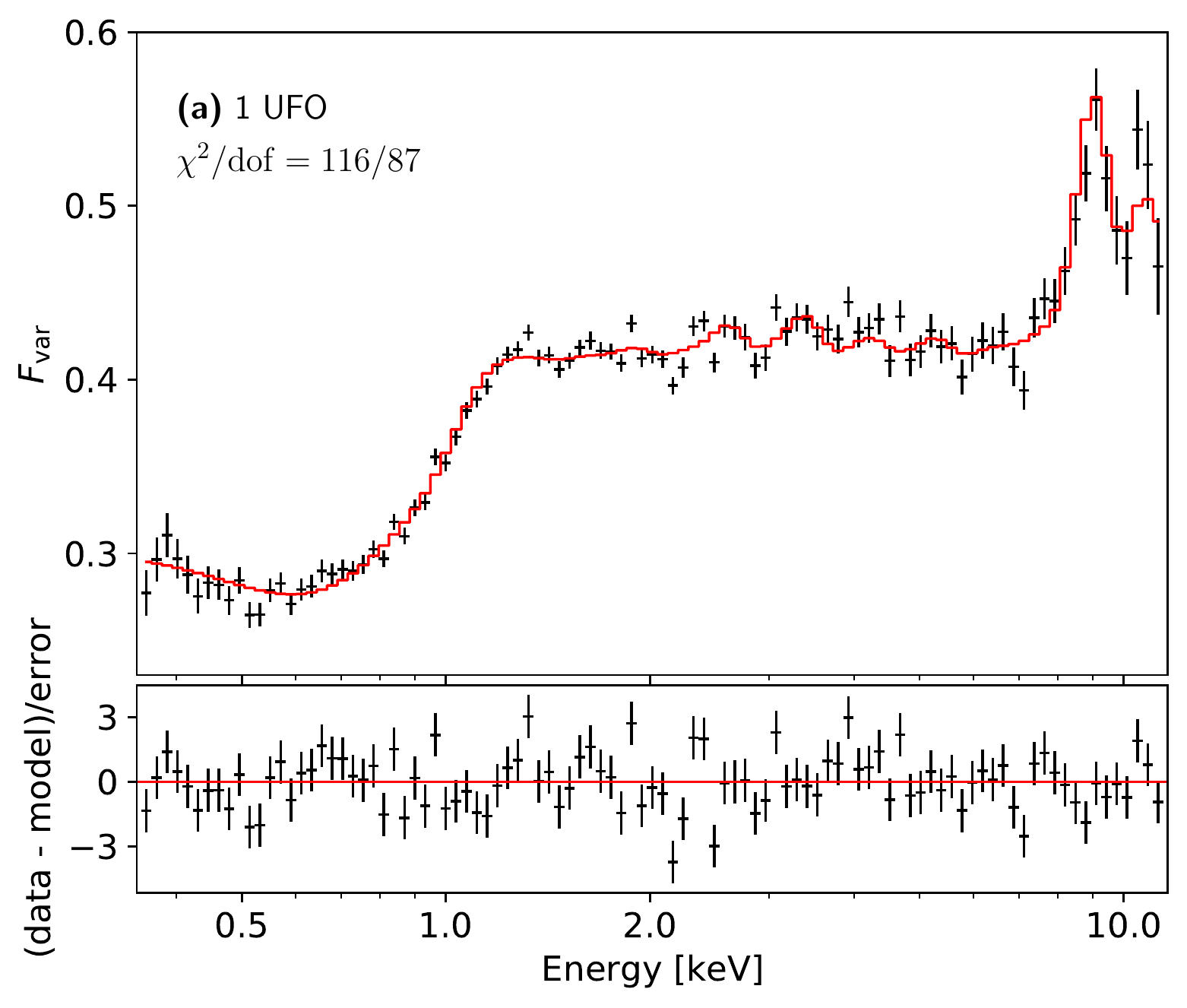}
    \includegraphics[width=0.45\linewidth]{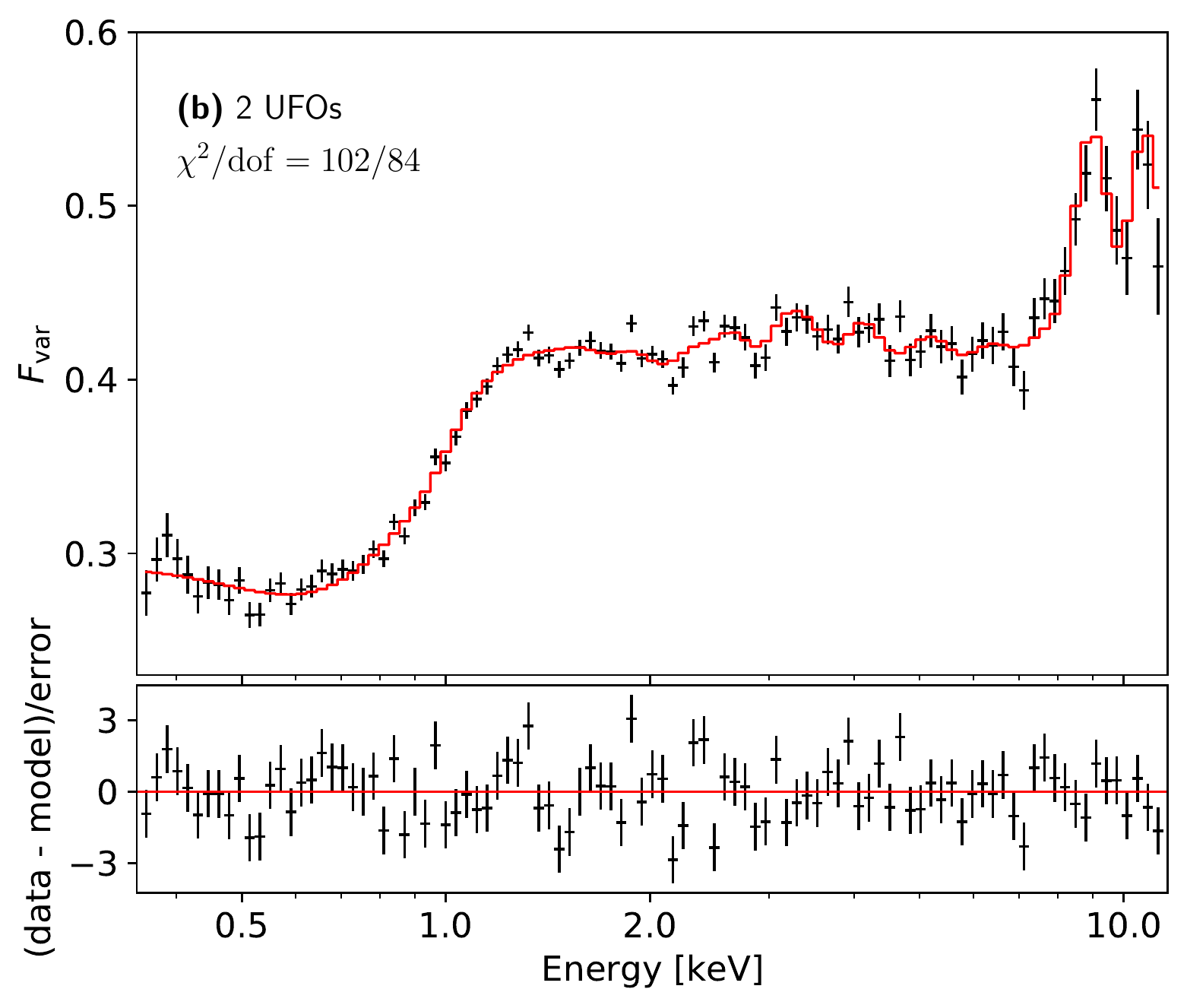}
    \includegraphics[width=0.45\linewidth]{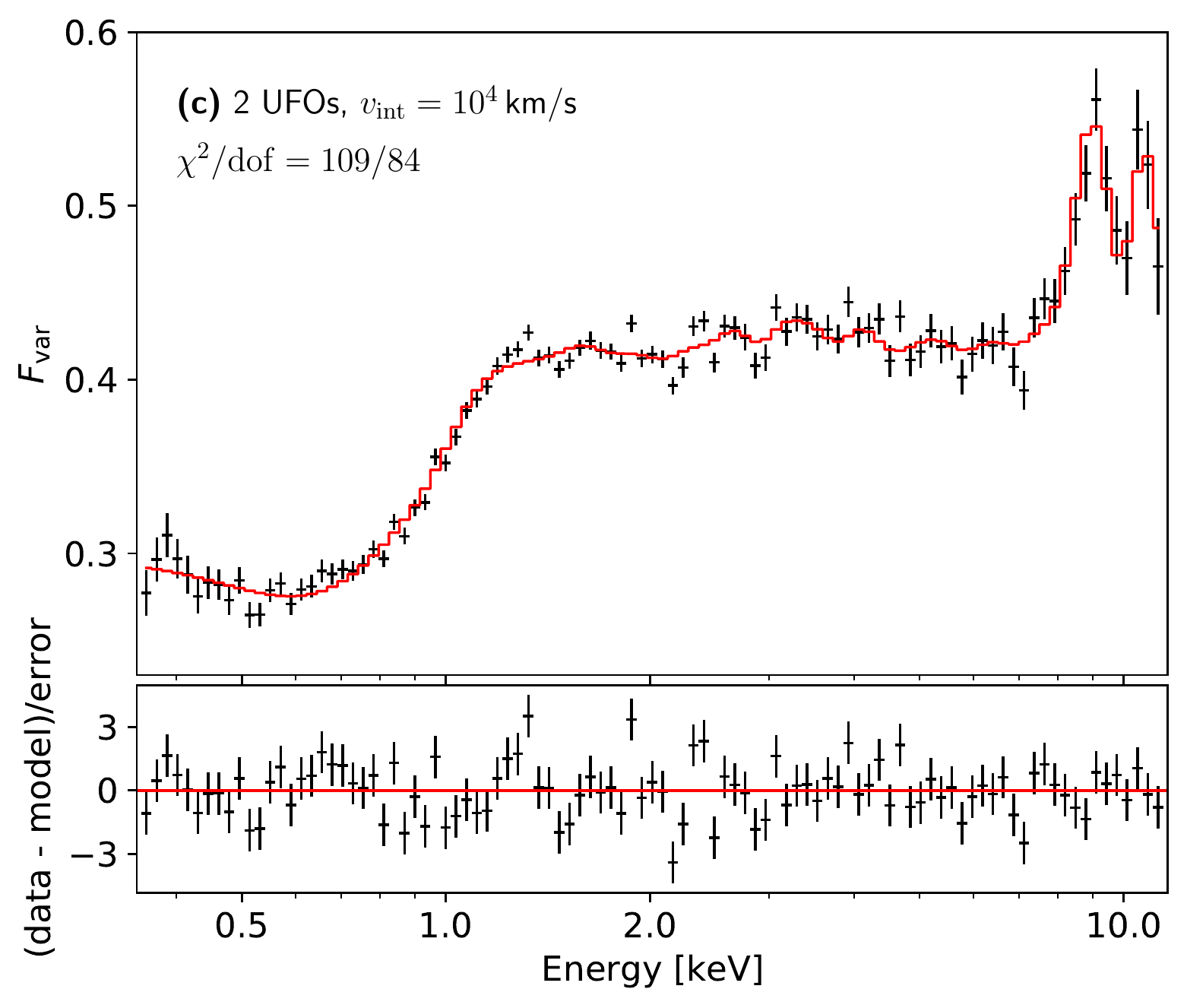}
    \includegraphics[width=0.45\linewidth]{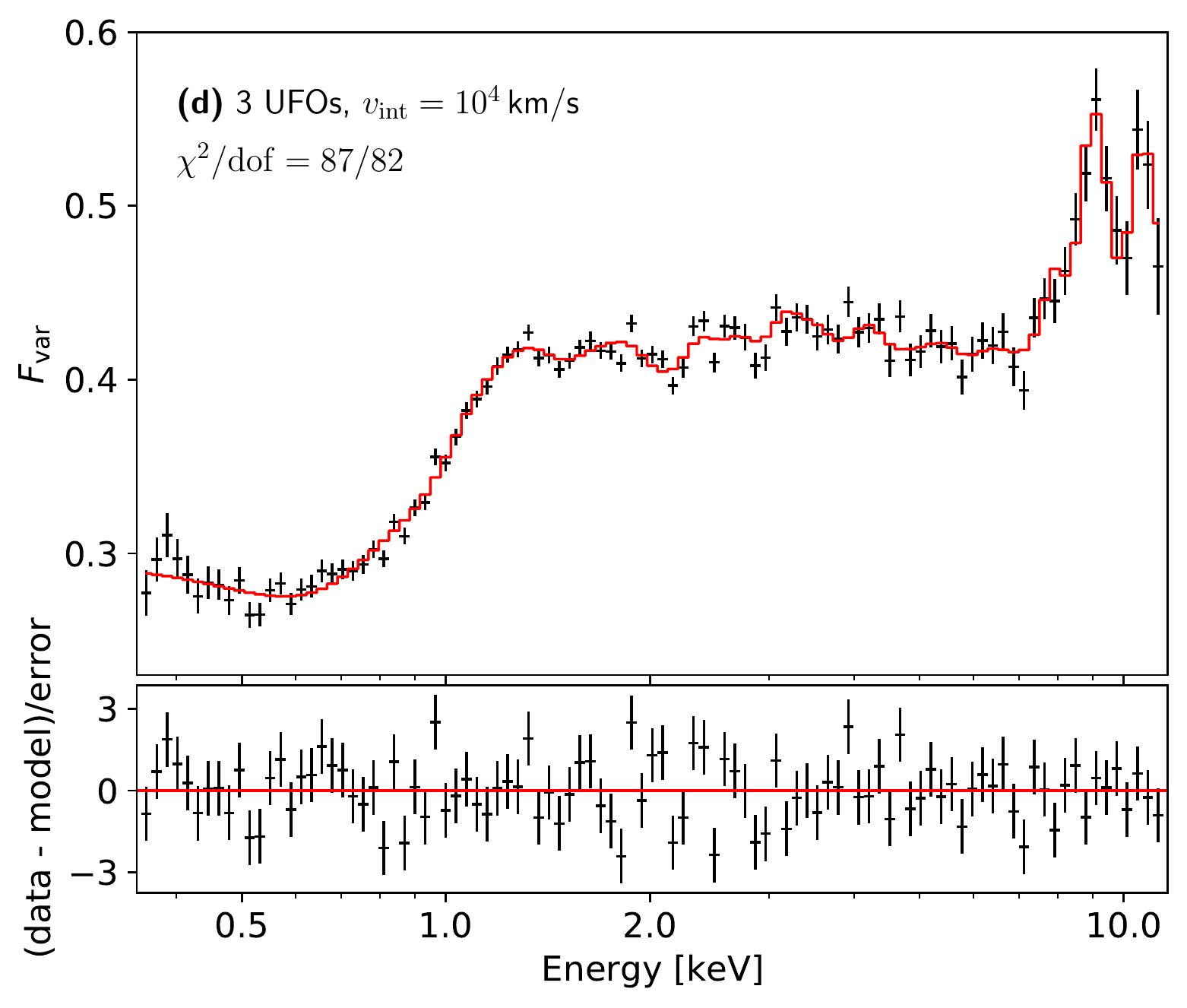}
    \caption{The RMS spectrum of PDS~456, modelled following \citet{Parker20}, with improvements to the UFO component to account for broadening of the UFO absorption spikes. To this end, we introduced a correlation between the outflow velocity and the logarithm of the flux and line broadening due to the intrinsic gas velocity, $v_\mathrm{int}$. Models (a) and (b) only include the first effect, while (c) and (d) assume an additional $v_\mathrm{int} = 10^4\,\mathrm{km}\,\mathrm{s}^{-1}$. One, two, and three outflowing layers are considered, in (a), (b) and (c), and (d) respectively. The continuum model includes power law variability, black body soft excess, as well as relativistic reflection. A systematic error of 0.01 was added. The energy is given in the reference frame of the source.}
    \label{fig:pds456}
\end{figure*}

The RMS spectrum of PDS~456 is flat above $\sim 1\,$keV, with some variation around the mean and two clear spikes just below and above $10\,$keV (see Fig.~\ref{fig:pds456}). At low energies, the variability is damped. We first construct a continuum model for this spectrum out of the components introduced by \citet{Parker20} and then add our improved UFO model. The continuum model includes two phenomenological components, a powerlaw (\textsc{fvar\_pow}) and blackbody soft excess (\textsc{fvar\_bbdamp}) to account for the damping, and relativistic reflection (\textsc{fvar\_refdamp}). For further details on these components, please refer to \citet{Parker20}. Second, we add the UFO model, with includes the improvements discussed in Sect.~\ref{sec:models}. Note that any constant multiplicative components, such as galactic absorption, can be ignored due to the fractional nature of RMS spectral models.

We start by considering a single UFO component without intrinsic Doppler broadening (\texttt{fvar\_ufo.fits}).
The result is shown in Fig.~\ref{fig:pds456} (a). Overall, the model provides a good description of the data ($\chi^2/\mathrm{dof} = 116/87$), keeping in mind that RMS spectral models are crude compared to count spectral models. The broadened Fe XXVI K$\alpha$ absorption spike is well accounted for by the improved UFO model, but the second spike is underestimated and some of the features between $1\mbox{--}3\,$keV are missed. The parameters of all fits are summarized in Tab.~\ref{tab:refpars}.

As discussed in the introduction, there is evidence that the UFO in PDS~456 consists of two or more layers with differences in outflow velocity and ionization. 
The second spike could therefore result from Fe XXVI K$\alpha$ absorption, as the first one, but from a layer with higher outflow velocity, i.e., larger blueshift. This interpretation is supported by the fact that both spikes have a similar intensity, which is hard to explain from a single velocity component, because Fe XXVI K$\alpha$ is expected to stand out clearly as the strongest line. We therefore add a second UFO component to our model, whereby the $C_\mathrm{UFO}$ and $C_v$ parameters are tied between components to reduce the number of free parameters. Adding a second UFO component does indeed significantly improve the fit ($\chi^2/\mathrm{d.o.f.} = 102/84$) and better account for the second spike (see Fig.~\ref{fig:pds456} (b)). However, the residuals in the $1\mbox{--}3\,$keV range remain largely the same.

Adding a further UFO component did neither decrease these residuals, nor improve the fit statistics. Therefore, we follow the second ansatz introduced in Sect.~\ref{sec:models}, i.e., to take into account line broadening due to the intrinsic velocity of the gas surrounding the AGN. Following estimates by \citet{Matzeu17} and \citet{Reeves16}, we choose a fixed $v_\mathrm{int}=10^4\,\mathrm{km}\,\mathrm{s}^{-1}$, which causes the broadening to be significant (see Fig.~\ref{fig:models_broad}), i.e., the value is well suited to investigate the influence of $v_\mathrm{int}$ on the UFO model.

The results are shown in Fig.~\ref{fig:pds456} (c). The fit is similar in quality to the one without intrinsic velocity broadening and there is no improvement in the $1\mbox{--}3\,$keV residuals. Most of the fit parameters (see Tab.~\ref{tab:refpars}) are equivalent within their error ranges, but the parameter which correlates the outflow velocity to the logarithm of the flux, $C_v$, is smaller. The smaller value is expected, as the velocity trend does not need to account for all the broadening which is observed, because $v_\mathrm{int}$ is included in the model.

To investigate the capabilities of the model with $v_\mathrm{int}$ further, we added a third component (Fig.~\ref{fig:pds456} (d)). As before, the correlation parameters $C_\mathrm{UFO}$ and $C_v$ were tied between components. In contrast to the previous result without intrinsic broadening, the additional third layer yields a very good fit ($\chi^2/\mathrm{dof} = 87/82$), which can be attributed to the fact that more of the features in the $1\mbox{--}3\,$keV range are now modelled correctly. However, this fit has a $C_v$ of the order of $10^{-8}$, i.e., the velocity trend is formally not required and all the broadening is due to intrinsic velocity broadening. In summary, both a velocity trend and intrinsic velocity broadening are independently able to account for the broadening of the absorption spikes, with a noticeable difference in the fit with three components. A detailed comparison of both effects will follow in Sect.~\ref{sec:discussion}. 

In general, it is difficult to constrain physical parameters given the current accuracy of the RMS spectral models, i.e., errors are large and often only a lower or upper bound can be obtained. Over all fits discussed above, the outflow velocity is $0.27\mbox{--}0.30\,c$ for the first and $0.41\mbox{--}0.49\,c$ for the second component. The third component in the last fit has a velocity of $0.15\mbox{--}0.20\,c$. These values agree well with previous studies \citep[e.g., ][]{Reeves18_pds456, Matzeu17, Boissay19, Reeves20}. All parameter ranges given here and in the following are 90$\,\%$ CL.

In the models without intrinsic velocity broadening, $C_v$ was found to be in the range $0.25\mbox{--}0.28$, which is a stronger correlation than reported by \citet{Matzeu17}, as shown in Fig.~\ref{fig:v-L} and discussed in detail in Sect.~\ref{sec:discussion}. If $v_\mathrm{int} = 10^4\,\mathrm{km}\,\mathrm{s}^{-1}$ is included, $C_v$ goes down to $0.22_{-0.08}^{+0.07}$, which means that the lower bound is consistent with \citet{Matzeu17}. For three layers and $v_\mathrm{int} = 10^4\,\mathrm{km}\,\mathrm{s}^{-1}$, the velocity trend is formally not required because $C_v$ is very small. The implications of these results will be discussed further in Sect.~\ref{sec:discussion}.

\begin{table*}
    \centering
    \caption{Fit parameters for PDS~456 resulting from the four different models shown in Fig.\ref{fig:pds456}. Further description of the \textsc{fvar\_ufo} component and its parameters can be found in the text. The continuum is described by a powerlaw component, \textsc{fvar\_pow}, blackbody soft excess, \textsc{fvar\_bbdamp}, and relativistic reflection, \textsc{fvar\_refdamp}, from \citet{Parker20}, where a detailed description of these components and their parameters can be found. The powerlaw parameters Var and $C_\Gamma$ describe the variance of the flux in log space and its correlation to the photon index, respectively. This correlation is very weak throughout the models, which is representative of the flat continuum variance. $C_\mathrm{ref}$ correlates the flux of the reflection component to the powerlaw flux.} Where the logarithm of a quantity is given, the unit refers to the quantity itself and not its logarithm.
    \begin{tabular}{l c l l l l l r}
    \hline
    \hline
    Component & Parameter & 1 UFO & 2 UFOs & 2 UFOs, $v_\mathrm{int}= 10^4\,\mathrm{km}\,\mathrm{s}^{-1}$ & 3 UFOs, $v_\mathrm{int}= 10^4\,\mathrm{km}\,\mathrm{s}^{-1}$ & Unit\\  
    \hline
    \textsc{fvar\_pow}      & Var             & $0.34_{-0.03}^{+0.02}$  & $0.34_{-0.03}^{+0.04}$& $0.35_{-0.03}^{+0.02}$& $0.31_{-0.03}^{+0.04}$    & \\
                            & $C_\Gamma$      & $\ll 0.01$              & $\ll 0.01$            & $\ll 0.01$            & $\ll 0.01$                & \\
    \textsc{fvar\_bbdamp}   & kT              & $0.31_{-0.02}^{+0.01}$  & $0.32 \pm 0.02$       & $0.32\pm 0.02$        & $0.31\pm 0.02$            & keV \\
                            & $f_\mathrm{BB}$ & $0.20 \pm 0.03$         & $0.18_{-0.03}^{+0.04}$& $0.20 \pm 0.03$       & $\geq 0.15$               & \\
    \textsc{fvar\_refdamp}  & $\log(n)$       & $\geq 18.9$             & $\geq 18.8$           & $\geq 18.8$           & $\geq 18.4$               & cm$^{-3}$ \\
                            & $\log(\xi)$     & $\leq 1.02$             & $\leq 1.09$           & $\leq 1.08$           & $\leq 1.1$                & erg~cm~s$^{-1}$  \\
                            & $f_\mathrm{ref}$&  $\geq 4.4$             & $\geq 3 $             & $\geq 4.4 $           & $\geq 3 $                 & \\
                            & $C_\mathrm{ref}$& $0.47_{-0.03}^{+0.04}$  & $0.49\pm 0.05$        & $0.47\pm 0.04$        & $0.56_{-0.07}^{+0.10}$    & \\
    \textsc{fvar\_ufo (1)}  & $\log(\xi)$     & $4.5_{-0.2}^{+0.1}$     & $4.10_{-0.12}^{+0.08}$& $4.1_{-0.1}^{+0.2}$   & $4.19_{-0.5}^{+0.3}$      & erg~cm~s$^{-1}$ \\
                            & $n_\mathrm{H}$  & $8.1_{-0.3}^{+0.2}$     & $1.7_{-0.4}^{+0.6}$   & $2.00_{-0.06}^{+0.13}$& $\leq 5.5$                & $10^{23}$\,cm$^{-2}$ \\ 
                            & $C_\mathrm{UFO}$& $\geq 0.7$              & $\geq 0.94$           & $\geq 0.4$            & $0.49_{-0.28}^{+0.08}$    & \\
                            & $C_v$           & $0.27_{-0.04}^{+0.09}$  & $0.26_{-0.05}^{+0.02}$& $0.22_{-0.08}^{+0.07}$& $\ll 0.01$                & \\
                            & $v$             & $-0.28 \pm 0.01$        & $ -0.28 \pm 0.01$     & $ -0.29 \pm 0.01$     & $ -0.28 \pm 0.01$         & $c$ \\
    \textsc{fvar\_ufo (2)}  & $\log(\xi)$     &                         & $4.1_{-0.1}^{+0.3}$   & $4.2_{-0.1}^{+0.4}$   & $4.0_{-0.2}^{+0.1}$       & erg~cm~s$^{-1}$ \\
                            & $n_\mathrm{H}$  &                         & $\leq 1.4$            & $\leq 4$              & $\leq 2.4$                & $10^{23}$\,cm$^{-2}$ \\
                            & $v$             &                         & $ -0.45_{-0.02}^{+0.04}$  & $ -0.43 \pm 0.02$ & $ -0.43 \pm 0.02 $        & $c$ \\
    \textsc{fvar\_ufo (3)}  & $\log(\xi)$     &                         &                       &                       & $3.77_{-0.04}^{+0.20}$    & erg~cm~s$^{-1}$ \\
                            & $n_\mathrm{H}$  &                         &                       &                       & $\leq 1.9$                & $10^{22}$\,cm$^{-2}$ \\
                            & $v$             &                         &                       &                       & $ -0.17^{+0.03}_{-0.02}$  & $c$ \\
    \hline
    \hline
    \end{tabular}
    \label{tab:refpars}
\end{table*}

\subsection{Revisiting IRAS~13224-3809}
\label{sec:IRAS}

IRAS~13224-3809 is a narrow line Seyfert 1 galaxy which is extremely variable on timescales as short as hours and therefore well suited to perform variability studies \citep[e.g.,][]{Parker17_nature, Alston19}. Here, we include our improved UFO model in the model of IRAS~13224-3809 introduced by \citet{Parker20}. All details concerning the data and modelling can be found in \citet{Parker20}.

We exclude all data below $3\,$keV, as our goal is to investigate the shape of the Fe~K$\alpha$ UFO spike. Therefore, it is sufficient to use the \textsc{fvar\_pow} model for the continuum, i.e., to ignore blackbody damping and relativistic reflection. As for PDS~456, a systematic error of 1\% was included in the analysis. Only broadening due to the correlation between outflow velocity and logarithm of the flux is considered, not intrinsic Doppler broadening.

\begin{figure}
    \centering
    \includegraphics[width=\linewidth]{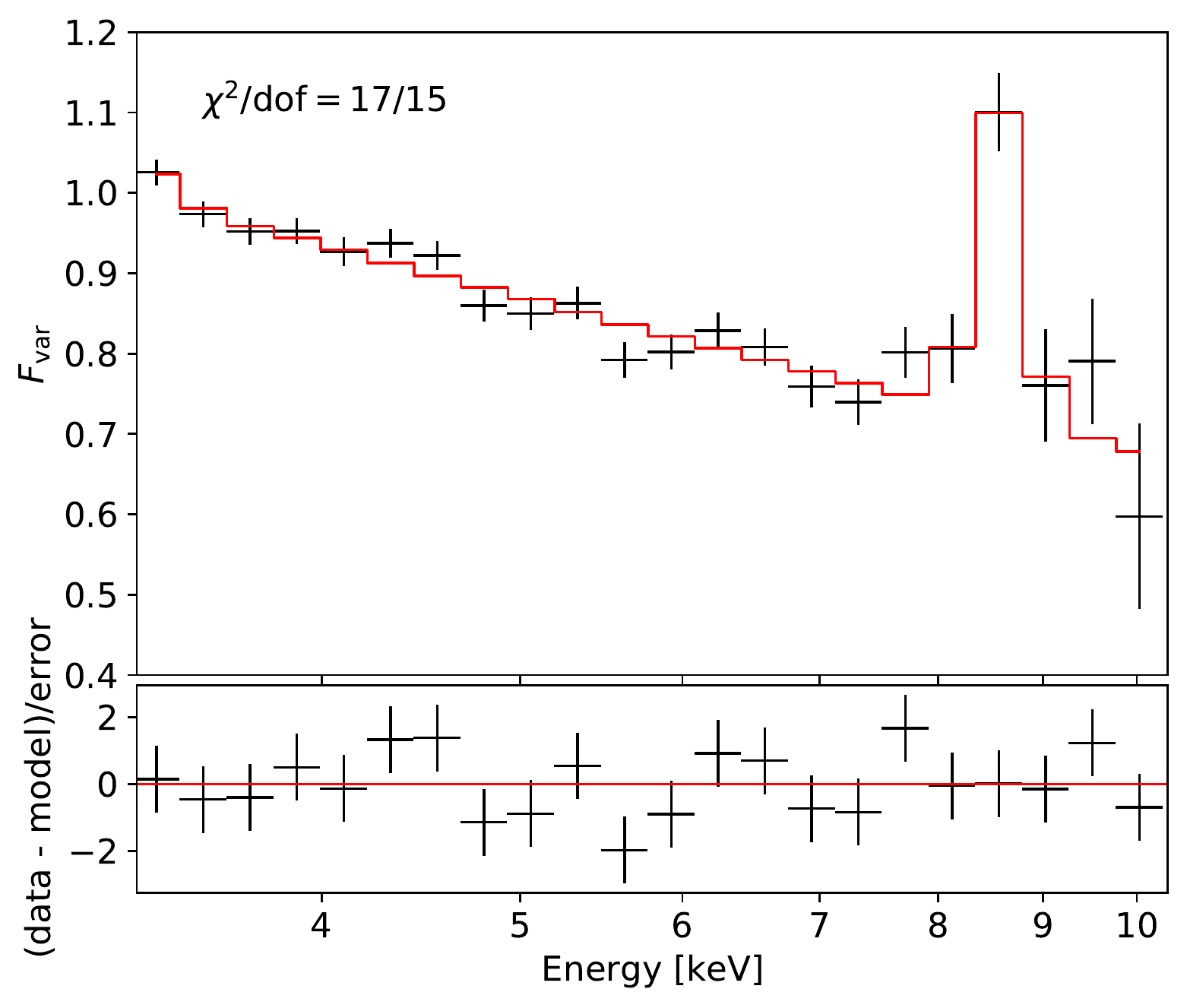}
    \caption{RMS spectrum of IRAS~13224 fitted with the improved UFO model, which accounts for broadening of the absorption spike by introducing a correlation between the outflow velocity and the logarithm of the flux (velocity trend). A 1\% systematic error was added. The fit has a similar quality as the previous model by \citet{Parker20} and places an upper limit of 0.02 on the correlation parameter $C_v$, which indicates that the velocity trend in IRAS~13224 is weak and not significant for understanding the RMS spectrum. The energy is given in the reference frame of the source.}
    \label{fig:iras13224_new-ufo}
\end{figure}

The fit result is shown in Fig.~\ref{fig:iras13224_new-ufo} and Tab.~\ref{tab:refpars_iras}. The spike shape is described well by the model. The parameters are generally consistent with the ones found by \citet{Parker20}. An upper bound of 0.02 is found on $C_v$, which suggests that the velocity trend is significantly smaller than for PDS~456, which is $C_v = 0.25\mbox{--}0.28$ in the model equivalent to the one used here, i.e., without intrinsic Doppler broadening. The smaller value is to be expected, as the spike in the RMS spectrum is noticeably narrower than for PDS~456. 

There is evidence for a small velocity trend in the literature \citep{Pinto18,Chartas18}. However, the precise strength is hard to determine, due to the strong variability of IRAS~13224-3809. In addition, the observed broadening does not necessarily originate from a velocity trend, but can also arise from other effects, e.g., intrinsic velocity broadening. Therefore, only upper bounds can be inferred from fitting models that account for a single effect. In the count spectrum, distinguishing line broadening and shift due to a velocity trend is easier. It is therefore promising to perform a joined analysis of RMS and count spectra in order to disentangle the different contributions.

\begin{table}
    \centering
    \caption{Fit parameters for IRAS~13224 obtained from modelling the $3\mbox{--}10\,$keV range of the RMS spectrum with the version of the improved UFO model, which excludes intrinsic velocity broadening (see Sect.~\ref{sec:models}). For this energy range, only the components \textsc{fvar\_ufo} and \textsc{fvar\_pow} need to be taken into account, as damping in the soft band is not important. Further description of \textsc{fvar\_ufo}  and \textsc{fvar\_pow} can be found in the text and \citet{Parker20} respectively. Where the logarithm of a quantity is given, the unit refers to the quantity itself and not its logarithm.}
    \begin{tabular}{l c l r}
    \hline
    \hline
    Component & Parameter & Value & Unit\\
    \hline
    \textsc{fvar\_pow}   &   Var & $0.44_{-0.05}^{+0.02}$  &\\
                        &   $C_\Gamma$  & $0.49_{-0.06}^{+0.05}$   &\\
    \textsc{fvar\_ufo}  &   $\log(\xi)$   & $4.1_{-0.5}^{+0.6}$  &erg~cm~s$^{-1}$\\
                        &      $n_\mathrm{H}$ & $\geq 10^{23}$ & cm$^{-2}$\\ 
                        &   $C_\mathrm{UFO}$& $0.013_{-0.006}^{+0.04}$  &\\
                        &   $C_v$& $\leq 0.02$  &\\
                        &   $v$ & $-0.23_{-0.01}^{+0.02}$  &    $c$\\
    \hline
    \hline
    \end{tabular}
    \label{tab:refpars_iras}
\end{table}


\section{Discussion}
\label{sec:discussion}

We have shown that the RMS spectrum of the luminous radio-quiet quasar PDS~456 is described well by the RMS spectral models from \citet{Parker20}, with improvements to the UFO component accounting for the observed broadening of the spikes in the RMS spectrum. Two effects were implemented to achieve this: a correlation between the outflow velocity and the logarithm of the X-ray flux via a parameter $C_v$, motivated by the discovery of such a velocity trend by \citet{Matzeu17}, and intrinsic Doppler broadening with a velocity $v_\mathrm{int}$. 

\subsection{The number of outflowing layers}

A single outflowing layer does not account for the similar height of the two prominent spikes in the RMS spectrum of PDS~456. Adding a second layer provides a satisfying result and significantly improves the fit quality, which we interpret as clear evidence that a second layer is present. This is the first time that strong evidence for a second outflowing layer has been found in EPIC-pn data alone, which speaks for the sensitivity of the method. Previous evidence for a second layer in PDS~456 was provided by a joined analysis of \nustar, \xmm\ \citep{Reeves18_pds456}, and \chandra\ data \citep{Boissay19}. Their measured outflow velocites of $0.46\pm0.02\,c$ and 0.48$\,c$ respectively are well within the range of our results ($0.41\mbox{--}0.49\,c$). 

A further significant improvement to the fit can be made by adding a third layer in the case where $v_\mathrm{int} = 10^4\,\mathrm{km}\,\mathrm{s}^{-1}$. However, the excellent quality of $\chi^2/\mathrm{dof} = 87/82$ is mostly because the features in the $1\mbox{--}3\,$keV range are described better, not the spikes, which are crucial to UFO detection in our approach. As this is a very good fit nonetheless, we interpret this as moderate evidence for a third layer. \citet{Reeves16} found a low ionization component of the wind at $0.1\mbox{--}0.2\,c$ in their analysis of data from RGS taken in 2013--2014, by fitting absorption lines of H- and He-like neon and L-shell iron below $\sim 2\,$keV. 
These results go well with our proposed third layer, i.e., 
the improved description in the $1\mbox{--}3\,$keV band and the slightly lower ionization ($3.77^{+0.20}_{-0.04}$) as compared to $4.19^{+0.3}_{-0.5}$ and $4.0^{+0.1}_{-0.1}$ for the first and second layer. The presence of line broadening in the count spectrum with $v_\mathrm{int}\sim 10^{4}\,\mathrm{km}\,\mathrm{s}^{-1}$ is well established for the first layer ($0.27\mbox{--}0.30\,c$) in the Fe~K band \citep[e.g.,][]{Matzeu17, Reeves20} and can likely serve as a valid approximation for the second ($0.41\mbox{--}0.49\,c$) and possible third ($0.15\mbox{--}0.20\,c$) layer for the purposes of our analysis \citep[e.g.,][]{Reeves16, Reeves20}. Considering the good fit result, it is therefore plausible that intrinsic Doppler broadening in the count spectrum contributes noticeably to spike broadening in the RMS spectrum in PDS~456. However, an intrinsic velocity is unlikely to be responsible for all broadening, as a velocity trend is known to exist in PDS~456 \citep{Matzeu17} and we have shown that such a trend causes broadening as well. 

In summary, our results are in agreement with the existing multi-layer picture of the UFO in PDS~456. That is, fast, highly ionized winds are launched close to the black hole, and are accompanied by a slower, possibly clumpy layer with low ionization further out. In addition, our method is able to provide evidence for multiple layers in a single analysis, because additional UFO components can be easily added to the fit, which is an advantage over count spectra analysis and helps piece together a unified picture. Fitting RMS spectra is therefore a promising approach to study the wind structure in PDS~456 and similar AGN. To allow for a more quantitative analysis, we aim to improve the accuracy of our model. This is of special interest regarding future micro calorimetry missions such as \xrism\ and \athena, which will provide X-ray spectroscopy data of unprecedented resolution, that is especially valuable in detecting UFO absorption in the soft band.

\subsection{Increase of overall variability by UFOs}

A general result of this kind of modelling is that the presence of UFOs enhances the observed X-ray variability of the AGN hosting them. In this case, the UFO components make up $5\mbox{--}8\,\%$ of the total $F_\mathrm{var}$ for the fits with one and two layers and $15\,\%$ for three. The corresponding value for IRAS~13224-3809, where more lines are present, is $\sim16$\,\%, suggesting that AGNs might have a $\sim 10\,\%$ enhanced variability if UFOs are present. 

Selecting variable sources for UFO searches, as in \citet{Igo20}, is therefore likely to be a promising way of detecting more UFOs. However, it is worth noting that such surveys may be biased towards a higher prevalence of UFOs because of this.


\subsection{Velocity trends vs. intrinsic line broadening as source for broadening of the UFO spikes}
\label{sec:broad}

\begin{figure}
    \centering
    \includegraphics[width=\linewidth]{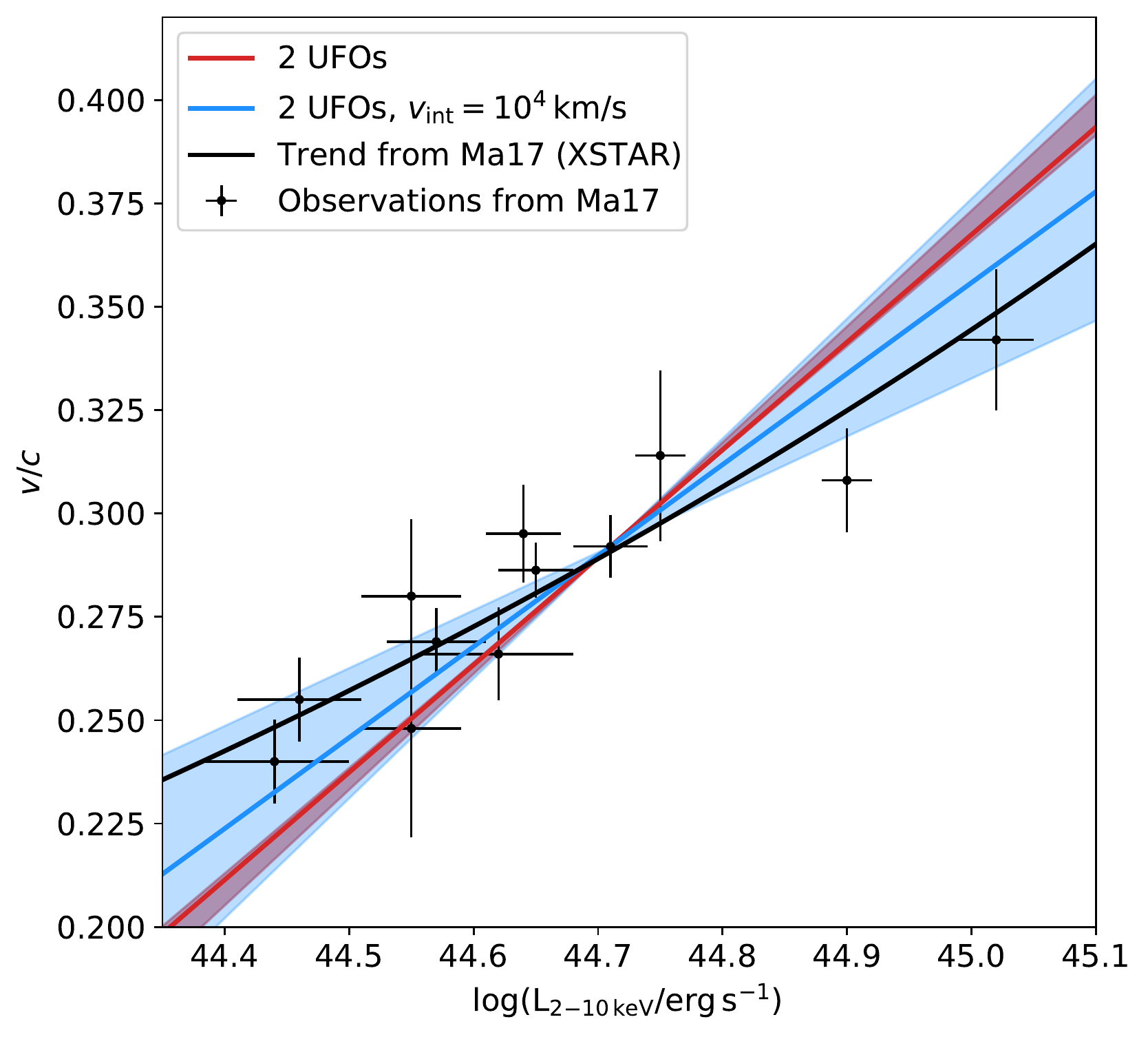}
    \caption{Measured relation (\textit{black points and regression line}) between the outflow velocity and the $2\mbox{--}10\,$keV luminosity of PDS~456 from \citet{Matzeu17} (Ma17 here), compared to the relations obtained from fitting the improved UFO model (\textit{red and blue}) to the RMS spectrum. The two models considered here assume two outflowing layers. One model includes line broadening due the intrinsic velocity of the gas (\textit{blue}) and one does not (\textit{red}), as in Fig.~\ref{fig:pds456} (b) and (c). The slope of the lines is given the the correlation parameters $C_v$ for these fits, with their 90\% CL shaded. The y offsets were chosen arbitrarily to allow for a by-eye comparison to the trend from Ma17. One sigma error bars are shown for the observations.}
    \label{fig:v-L}
\end{figure}

The broadening of the UFO spikes can be modelled equally well by a velocity trend and intrinsic velocity broadening, which begs the question of how their individual contributions can be disentangled. Figure~\ref{fig:v-L} compares $C_v$ to the velocity trend found by \cite{Matzeu17}, for fits with and without velocity broadening. It can be seen that the fit without $v_\mathrm{int}$ most likely overestimates the trend and does not agree well with the data, while the case of $v_\mathrm{int} = 10^4\,\mathrm{km}\,\mathrm{s}^{-1}$ is consistent with the lower limit of the range. In conclusion, $v_\mathrm{int}$ contributes to the spike broadening in PDS~456 and $C_v$ can only be interpreted as an upper limit on the strength of the velocity trend if $v_\mathrm{int}$ is ignored. 

As already mentioned in Sect.~\ref{sec:IRAS}, separating both effects from RMS spectra alone is difficult. Performing an additional analysis of a single-epoch count spectrum is a promising approach to solve this problem, because line broadening and shift due to a velocity trend can be more clearly distinguished in count spectra. However, for more rapidly variable sources like IRAS~13224-3809, a count spectrum with sufficient signal to detect the UFO lines will integrate over a wide range of fluxes and hence broadening will be introduced as a result of the velocity trend, i.e., the line shift with changing flux. In this case, a flux-resolved analysis is likely the most effective way of identifying the strength of the velocity trend \citep[see][]{Pinto18}.


\section{Conclusions}

Modelling RMS spectra is a promising approach to understand the variability of AGN. We improve upon a model by \citet{Parker20} for UFO absorption, that manifests itself through spikes in the variance. We do this by including two effects which cause these spikes to broaden as seen in the RMS spectrum of PDS~456, i.e., a correlation between the outflow velocity and the logarithm of the X-ray flux and intrinsic Doppler broadening. 

Both effects are able to describe the broadened spikes in PDS~456 equally well. We find clear evidence for two outflowing layers and possible indication of a third one. This is the first time a fast layer ($0.41\mbox{--}0.49\,c$) has been seen in \xmm\ EPIC-pn data alone, which highlights the potential of using RMS spectral models to detect UFOs and investigate the wind structure. Our results agree very well with the existing notion of a multi-layered UFO in PDS~456 and with previous measurements of the outflow velocities. If intrinsic Doppler broadening is excluded, the value for the strength of the correlation between the outflow velocity and the logarithm of the X-ray flux is overestimated, suggesting that Doppler broadening is needed and if excluding it, the correlation strength has to be interpreted as an upper limit.

We briefly revisit IRAS~13224-3809 and find that its upper limit on the correlation strength is smaller than for PDS~456, as expected from the narrower spike in its RMS spectrum. From modeling RMS spectra alone, it is difficult to disentangle the different effects contributing to the broadening. We therefore suggest the analysis of RMS and count spectra to go hand in hand in future studies.

We aim to further improve the UFO model and expand the set of existing models as modelling RMS spectra is very likely to provide novel insights into the variability of AGN.


\section*{Acknowledgements}

We thank the anonymous referee for their constructive feedback. MLP and GM are supported by European Space Agency (ESA) Research Fellowships. Based on observations obtained with XMM-Newton, an ESA science mission with instruments and contributions directly funded by ESA Member States and NASA. 

\section*{Data Availability}

All data used in this work are publicly available from the \xmm\ science archive (\texttt{nxsa.esac.esa.int}). The models are publicly available from \texttt{www.michaelparker.space/xspec-models}.




\bibliographystyle{mnras}
\bibliography{bibliography} 




\bsp	
\label{lastpage}
\end{document}